\documentclass[preprintnumbers,amsmath,amssymb,nofootinbib,superscriptaddress,floatfix,showkeys,twocolumn]{revtex4-1}
\usepackage{graphicx,color}
\usepackage[latin1]{inputenc}
\usepackage{amsmath,amssymb}
\usepackage{hyperref}
\usepackage{epstopdf}
\usepackage{slashed}
\usepackage{soul}
\usepackage{amssymb}
\usepackage[normalem]{ulem}
\usepackage[caption=false]{subfig}

\newcommand{\mnras}{Mon. Not. R. Astron. Soc.}

\newcommand{\aap}{Astron. Astrophys}
\newcommand{\aj}{Astron. J.}

\newcommand{\beq}{\begin{equation}}
\newcommand{\eeq}{\end{equation}}
\newcommand{\beqa}{\begin{eqnarray}}
\newcommand{\eeqa}{\end{eqnarray}}

\newcommand{\apjl}{Astrophys. J.}

\begin{document}

\title{Evidence for GeV emission of the superluminous supernova SN 2017egm}

\author{Shang Li}\email{Contributed equally}
\affiliation{School of Physics, Anhui University, Hefei 230601, China}
 
\author{Yun-Feng Liang}\email{Contributed equally}
 
\affiliation{Guangxi Key Laboratory for Relativistic Astrophysics, School of Physical Science and Technology, Guangxi University, Nanning 530004, China}

\author{Neng-Hui Liao}
\affiliation{Department of Physics and Astronomy, College of Physics, Guizhou University, Guiyang 550025, China}
\author{Lei Lei}
\affiliation{Key Laboratory of Dark Matter and Space Astronomy, Purple Mountain Observatory, Chinese Academy of Sciences, Nanjing 210023, China}
\affiliation{School of Astronomy and Space Science, University of Science and Technology of China, Hefei 230026, China}
\author{Yi-Zhong~Fan}
\email[]{Corresponding author: yzfan@pmo.ac.cn}
\affiliation{Key Laboratory of Dark Matter and Space Astronomy, Purple Mountain Observatory, Chinese Academy of Sciences, Nanjing 210023, China}
\affiliation{School of Astronomy and Space Science, University of Science and Technology of China, Hefei 230026, China}


\begin{abstract}
Superluminous supernovae (SLSNe) are a new class of transients with luminosities $\sim10 -100$ times larger than the usual core-collapse supernovae (SNe). Their origin is still unclear and one widely discussed scenario involves a millisecond magnetar central engine. The GeV-TeV emission of SLSNe has been predicted in the literature but has not been convincingly detected yet. Here we report the results of the search for $\gamma$-ray emission in the direction of SN 2017egm, one of the closest SLSNe detected so far, using 15 years of {\it Fermi}-LAT Pass 8 data. There is a transient $\gamma$-ray source appearing about 2 months after this event and lasting a few months. Monte Carlo simulations show that the $\gamma$-ray signal has a global significance of {\it at least} 4$\sigma$. Both the peak time and the luminosity of the GeV emission are consistent with the magnetar model prediction, suggesting that such a GeV transient is the high-energy counterpart of SN 2017egm and the central engine of this SLSNe is a young magnetar. 

\end{abstract}

\maketitle

\section{\label{sec::introduction}Introduction}

Superluminous supernovae (SLSNe) are a new type of supernovae and their luminosities are $\sim$10$-$100 times larger than usual core-collapse supernovae (SNe)
\cite{2011Natur.474..487Q,2012Sci...337..927G,2019ARA&A..57..305G}. They are so bright that can be detected in optical at redshifts of $\geq 3.5$ \cite{2012Natur.491..228C}. 
The energy sources of these energetic explosions are still in debate and the widely discussed scenarios include $^{56}$Ni decay, the interaction between ejecta and dense circumstellar media \cite{Moriya:2018sig}, and the spin-down of the nascent magnetar \cite{2007ApJ...666.1069M,2010ApJ...717..245K,2010ApJ...719L.204W,Murase:2021lro}.  
Theoretically, fast particle acceleration might occur soon after the supernova explosion, supposing the presence of interaction between the ejecta and pre-existing dense material \cite{2011PhRvD..84d3003M} or the existence of a nascent nebula \cite{2015ApJ...805...82M}. Therefore, SLSNe can be a new kind of GeV-TeV $\gamma$-ray source. 

GeV Observations provide a probe for the central engines of SLSNe, crucial for solving the fundamental question of their energy source.
Dedicated efforts have been made to search for $\gamma$-ray emission from SLSNe with the {\it Fermi} Large Area Telescope ({\it Fermi}-LAT) \cite{Atwood_2009} observation data, including both individual analysis and stacking procedures, but no reliable signal has been detected so far \cite{2018A&A...611A..45R,2023arXiv230206686A}. A variable $\gamma$-ray source, with an isotropic energy of $\sim 10^{51}~{\rm erg}~(D_{\rm L}/156~{\rm Mpc})^{2}$ and a duration of $\sim $2 years, was detected in the direction of a peculiar supernova iPTF14hls \cite{Yuan_2018,Prokhorov:2021tkz}, where $D_{\rm L}$ is the luminosity distance of the source. If this $\gamma$-ray signal was from iPTF14hls, the GeV emission efficiency should be very high. However, a quasi-stellar object SDSS J092054.04+504251.5, which is likely a blazar according to the infrared data, is found in the error circle of this GeV source. Therefore, it is unclear whether the supernova iPTF14hls is indeed a $\gamma$-ray emitter. Below we focus on the SLSNe SN 2017egm for two good reasons. One is that this event is one of the closest hydrogen-poor (Type-I) SLSNe at a redshift of $z=0.0307$ ($D_{\rm L}=135~{\rm Mpc}$) \cite{2017ApJ...845L...8N}. The other is that the $\gamma$-ray emission with a peak luminosity of $\sim 10^{43}~{\rm erg~s^{-1}}$ appearing at $t\sim 3$ months after the stellar explosion has been predicted in the modeling of its optical emission within the pulsar wind nebula scenario \cite{Vurm_2021}, which should have been detected by {\it Fermi}-LAT if it is indeed the case. 
 
SN 2017egm was discovered by the $Gaia$ satellite on 2017 May 23 \cite{2017TNSTR.591....1D}. 
The radio observation of SN 2017egm was carried out after the explosion, some constraints were obtained \cite{2018ApJ...853...57B}. 
The X-ray emission of SN 2017egm was not detected by the four-epoch Chandra observations, leading to a constraint on the luminosity with a 3$\sigma$ upper limit of 1.9 $\times 10^{39}\rm  erg \rm~ s^{-1}$ in the 0.5 $-$ 10 keV range \cite{Zhu:2023ntt}.
Ref.~\cite{Zhu:2023ntt} also analyzed the high-cadence ultra-violet, optical, and near-infrared light curves of SN 2017egm and suggested that the sharp peak in the light curve and the multiple post-peak bumps can only be explained by the interactions between the ejecta and circumstellar material. 
However, hydrogen-poor SLSNe like SN2017egm are generally thought to be powered by a central engine. Under the central engine model, high-energy $\gamma$-rays generated in the pulsar/magnetar wind nebula are predicted to escape the SN ejecta months after the optical peak \cite{Vurm_2021}. Gamma-ray observations would thus provide crucial information for establishing the energy source of SN2017egm.
The $\gamma$-ray emission from SN 2017egm has been searched for in a wide energy range using the VERITAS and Fermi-LAT data \cite{2023arXiv230206686A}. VERITAS did not detect a significant signal, which can be easily explained by the absorption of very-high-energy $\gamma$ rays by the extragalactic background light. The most significant signal is found in the first 6 month period with TS $\sim$ 10.1 by {\it Fermi}-LAT\footnote{Note that Ref.~\cite{2023arXiv230206686A} used a 6 months time period and an energy range from 100 MeV to 500 GeV in their Fermi-LAT analysis, which is different from those used in this work. This can explain the difference in the obtained TS value.}. 
This indicates that a hint for detection had already appeared in the previous search. In this work, we carry out a dedicated analysis of the 15 years of {\it Fermi}-LAT data, and report a promising $\gamma$-ray counterpart of SN 2017egm along with some discussions and conclusions.

\begin{figure*}
\centering
\includegraphics[width=0.7\textwidth]{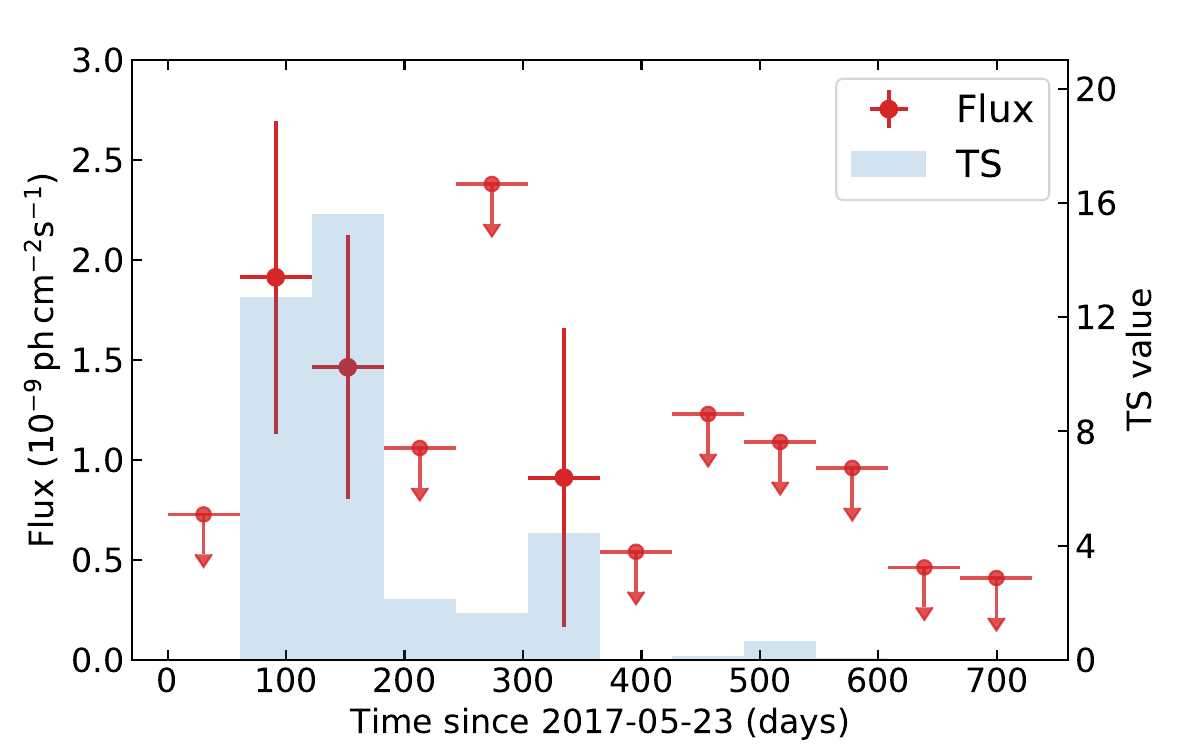}
\caption{Light curve of the $\gamma$-ray emission from the direction of SN 2017egm between 500 MeV and 500 GeV in 2-month bins. The zero point is set to the onset of the supernova, 2017 May 23. Shaded regions show the TS values (right axis). For TS values less than 4, 95\% confidence level upper limits are presented.}
\label{fig:lc}
\end{figure*}

\section{data Analysis} \label{sec:data}
We use 15 years of {\it Fermi}-LAT Pass 8 data (from August 4, 2008, to August 4, 2023) and the {\tt Fermitools} software to perform the analysis. The {\tt evtype=3}, {\tt evclass=128} events and the instrument response functions (IRFs) of {\tt P8R3\_SOURCE\_V3} are employed. Only photons in the energy range from 500 MeV to 500 GeV are selected. We exclude photons below 500 MeV because the expected spectrum of the SLSNe is relatively hard \cite{2015ApJ...805...82M}. Moreover, the poor angular resolution at low energies could cause contamination from the bright nearby source 4FGL J1015.0+4926, located a few degrees away. The maximum zenith angle ($Z_{\rm max}$) is set to $100^\circ$ to reduce the contamination from the Earth Limb. In addition, the quality-filter cuts ({\tt DATA\_QUAL==1 \&\& LAT\_CONFIG==1}) are applied to exclude the bad time intervals not suitable for science analysis. The standard binned likelihood analysis implemented with the {\tt gtlike} tool is used. The region-of-interest (ROI) is taken to be a 14$^{\circ}$ $\times$ 14$^{\circ}$ box centered at the target.

The script {\tt make4FGLxml.py}\footnote{\url{https://fermi.gsfc.nasa.gov/ssc/data/analysis/user/make4FGLxml.py}} is applied to generate the initial background model, in which all 4FGL-DR4 \cite{Ballet:2023qzs} sources within 15$^{\circ}$ around the target together with the two diffuse backgrounds ({\tt gll\_iem\_v07.fits} and {\tt iso\_P8R3\_SOURCE\_V3\_v1.txt}) are included. Due to the absence of SN 2017egm in the current $\gamma$-ray catalog, we add a new source at the optical position of SN 2017egm with a power-law spectrum (i.e. $dN/dE \propto E^{-\Gamma}$, where $\Gamma$ is the spectral photon index) into the model file to model the supernova component. In the data analysis, the parameters of the 4FGL-DR4 sources within a 7$^{\circ}$ radius centered at the target as well as the normalization parameters of the two diffuse components are left free, while others are fixed to their 4FGL-DR4 values. The test statistic (TS = 2$\rm \Delta$log$\mathcal{L}$) \cite{1996ApJ...461..396M} is adopted to quantify the significance of a $\gamma$-ray source, where $\mathcal{L}$ represents maximum likelihood value and $\Delta$log$\mathcal{L}$ is the log-likelihood difference between models with and without the new source. 

\section{results}\label{sec:results}

\begin{figure*}
\centering
\includegraphics[width=0.7\textwidth]{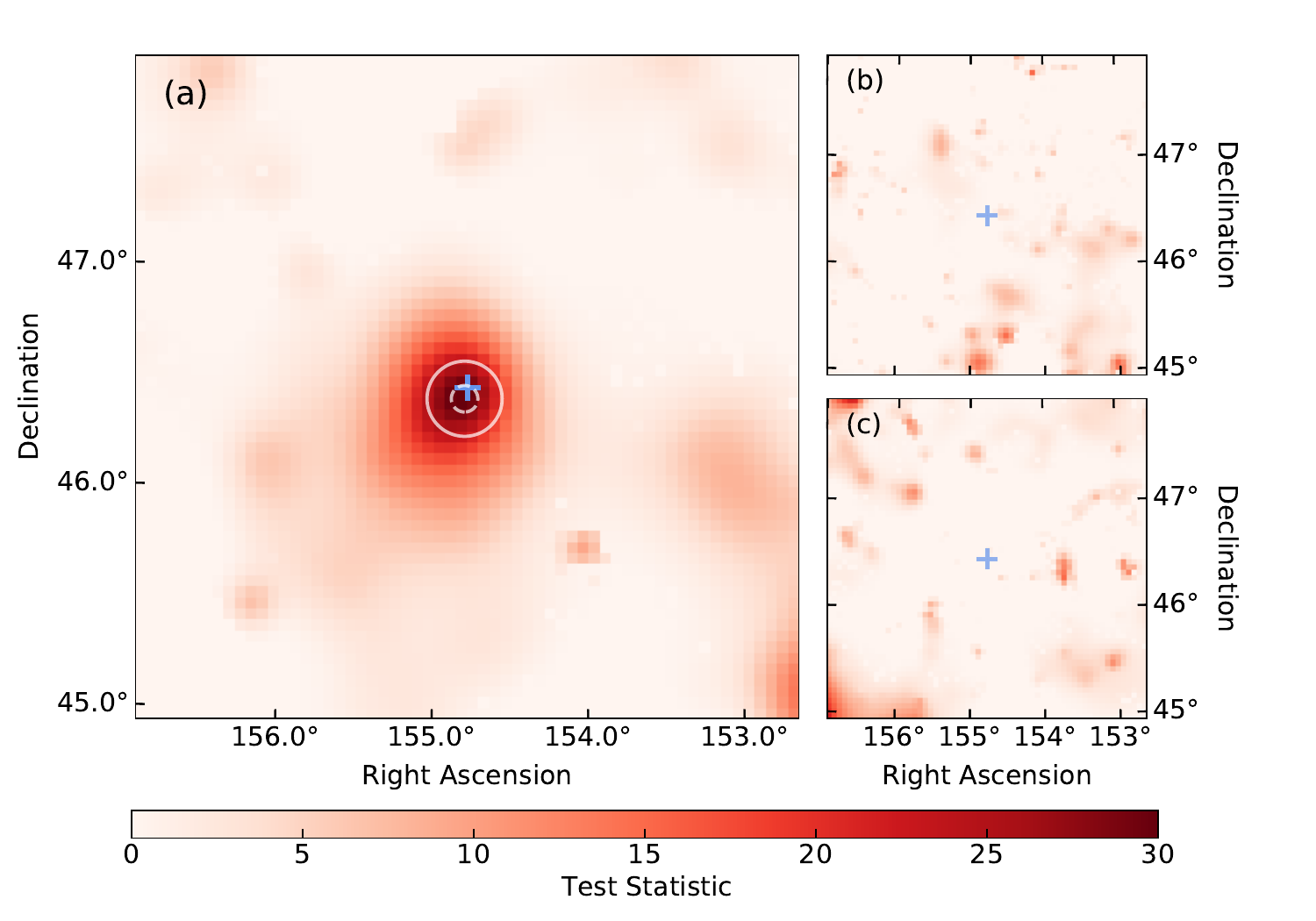}
 \caption{TS maps for different epochs: (a) from 2017 July 23 to 2017 November 23, (b) from 2008 August 4 to 2017 July 23, and (c) from 2017 November 23 to 2023 August 4. The TS maps display $3^{\circ}\times 3^{\circ}$ region centered at SN 2017egm for the data between 500 MeV and 500 GeV. The optical position of SN 2017egm is represented by the blue cross symbol. The circles in panel (a) are the 68\% (inner) and 95\% (outer) error uncertainties of the $\gamma$-ray source localization.}
\label{fig:tsmap} 
\end{figure*}

Initially, we fit the entire 15-year data and generate a residual TS map. Any excess in the TS map with a peak TS value of $>25$ is treated as a new point source and will be added to the background model with a power-law spectrum. Their positions are determined using the {\it gtfindsrc} tool. The TS map reveals two new $\gamma$-ray sources (TS $>$ 25) with TS values of 31 and 27. Their best-fit positions are $({\rm R.A.},{\rm Dec.})=(157.81^{\circ},45.21^{\circ})$ and $({\rm R.A.},{\rm Dec.})=(154.12^{\circ},44.46^{\circ})$, respectively. After adding the two sources to the background model, the likelihood fitting is performed once again to search for the emission from SN 2017egm. No significant $\gamma$-ray excess is found in the direction of SN 2017egm (TS $\sim$ 0.7 with $\Gamma$ fixed at 2.0) based on the 15-year data. 
However, theoretical investigations predict that the $\gamma$-ray emission lasts only weeks to months, depending on the characteristics of the SNe and the central compact object \cite{2015ApJ...805...82M}. Therefore, we perform another fit with the two-year data (from May 23 2017 to May 23 2019) after the explosion, which results in a TS value of $\sim$ 15.8 for SN 2017egm. 

Furthermore, we derive a 2-month-bin light curve of the $\gamma$-ray emission in the direction of SN 2017egm, which is shown in Fig.~\ref{fig:lc}. In the figure, for the bins with TS $<$ 4, 95\% confidential level (C.L.) upper limits are calculated using {\it pyLikelihood UpperLimits} tool. As shown in Fig.~\ref{fig:lc}, the TS values are small ($<$ 4) in most time bins. However, for the second and third bins, the TS values are larger than 12. These two bins correspond to the time range from 2017 July 23 to 2017 November 23. The following analysis is conducted for this time interval ($\sim$ 4-month). During this time period, analysis of SN 2017egm (using its optical position) detects a relatively strong $\gamma$-ray signal with a TS value of 29.6 (see the (a) panel of Fig.~\ref{fig:tsmap}). A localization analysis is performed for the new $\gamma$-ray source and the optimized coordinates are ${\rm R.A.}=154.79^{\circ}$ and ${\rm Dec.}=46.40^{\circ}$ with a 95\% C.L. error radius of 0.18${^\circ}$. The SN 2017egm is only 0.06$^{\circ}$ away from the best-fit position of the $\gamma$-ray source. Using the new $\gamma$-ray position, the TS value is $\sim$ 30.1 and the corresponding $\gamma$-ray flux is $\rm (1.9 \pm 0.6) \times 10^{-9}$ ph $\rm cm^{-2}$ $\rm s^{-1}$ between 500 MeV and 500 GeV. If the $\gamma$-ray source is indeed related to SN 2017egm, the averaged isotropic $\gamma$-ray luminosity in the period is $(9.0 \pm 2.4) \times 10^{42} (D_{\rm L}/135~{\rm Mpc})^{2}$ erg $\rm s^{-1}$. The spectral energy distribution of the $\gamma$-ray source is presented in Fig.~\ref{fig:sed} of the Supplementary Materials.

Furthermore, the data is divided into three parts (from 2008 August 4 to 2017 July 23, from 2017 July 23 to 2017 November 23 and from 2017 November 23 to 2023 August 4) to perform the likelihood analysis according to the new position. For the first and the third parts, no significant $\gamma$-ray signal appears at the position of SN 2017egm (see the (b) and (c) panels of Fig.~\ref{fig:tsmap}). However, a bright $\gamma$-ray signal (TS $\sim$ 30.0) appears in the direction of SN 2017egm in the second period (see the (a) panel of Fig.~\ref{fig:tsmap}), confirming that found in the 2-month bin $\gamma$-ray light curve.

\begin{figure*}[!ht]
\begin{centering}
\includegraphics[width=0.7\textwidth]{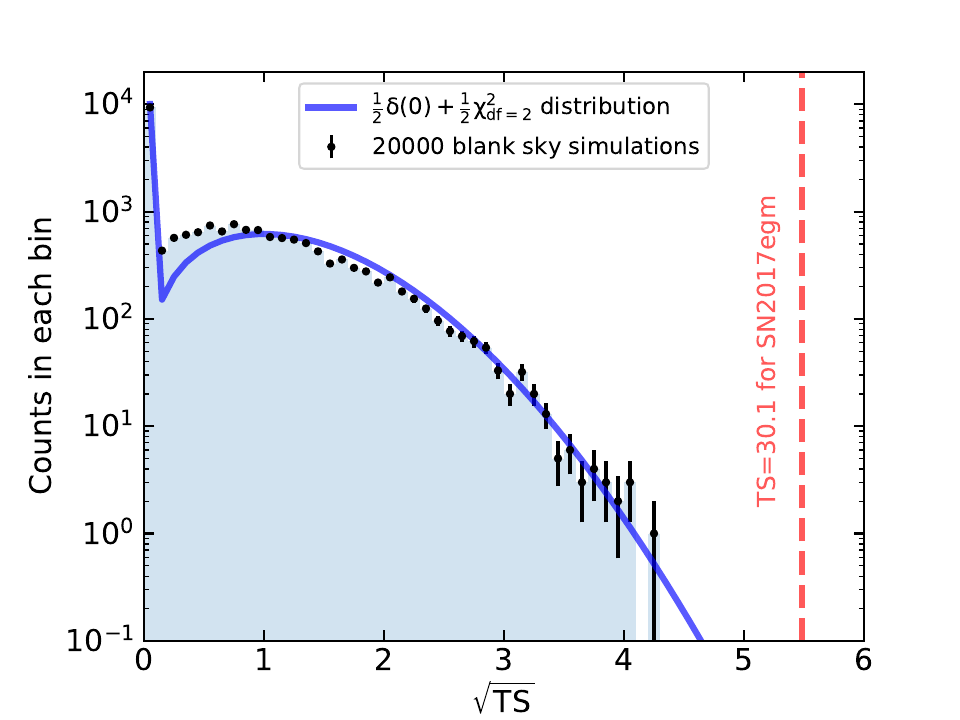}
\end{centering}
\caption{$\rm \sqrt{TS}$ distribution of 20000 random sky simulations. The blue curve is the TS distribution expected from Chernoff's theorem \cite{Chernoff1954}.}
\label{fig:S1}
\end{figure*}

We also performed random sky simulations to estimate the detection significance of the $\gamma$-ray signal. Considering the SN 2017egm is located at high galactic latitude ($|b| > 50^{\circ}$) and the nearest point source in 4FGL-DR4 to SN 2017egm is $\sim$ 2$^{\circ}$, the blank-sky locations are randomly generated at high Galactic latitude ($|b| > 30^{\circ}$) and stay away from any 4FGL-DR4 catalog sources (no point sources are within $1^{\circ}$ and no spatially extended sources are within $5^{\circ}$). For each blank-sky position, the corresponding time intervals are randomly generated within the time range from 2008 August 4 to 2023 August 4 for a 4-month period. In total, 20000 data sets are generated and processed with the same data cut and energy range as the real analysis. For each position, a corresponding point source with a single power-law spectrum is added into the model and the binned likelihood analysis is performed. The distribution of $\rm \sqrt{TS}$ is shown in Fig.~\ref{fig:S1}. 
Due to the maximum TS value of the 20000 simulations is less than 25, this implies that the probability of TS $\sim$ 30.1 occurring due to fluctuation is less than 1/20000, i.e., a p-value of $<5\times 10^{-5}$. Therefore, the detection significance of the $\gamma$-ray signal (TS $\sim$ 30.1) is at least $4 \sigma$.
The blue line in Fig.~\ref{fig:S1} represents the TS distribution expected from Chernoff's theorem \cite{Chernoff1954}, which roughly agrees with our simulation results. Using this theoretical distribution for the calculation, the signal significance is 5.3$\sigma$.

\begin{figure*}
\centering
\includegraphics[width=0.7\textwidth]{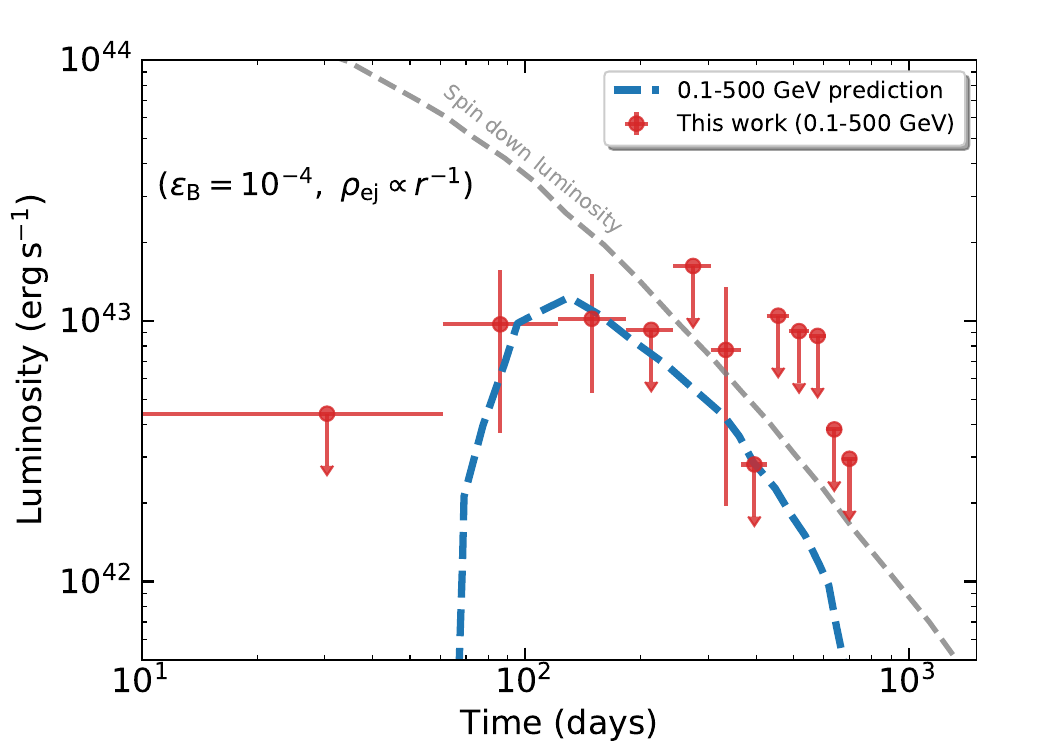}
\caption{
A comparison between the theoretical prediction (blue dashed line) and the observed $\gamma$-ray light curve (red points, in 2-month bin) of SN 2017egm. The theoretical curve considers a magnetar-powered supernova model \cite{Vurm_2021}. The nebula magnetization is assumed to be $\varepsilon_{B} = 10^{-4}$.
The dashed line shows the spin-down luminosity of the central magnetar.
}
\label{fig:fermi}
\end{figure*}

\section{discussion and conclusion}

It is possible that the $\gamma$-ray excess we detected in the direction of SN 2017egm, even though not a statistical fluctuation, might originate from another astrophysical source that happens to coincide with the position of the supernova. To examine this possibility, we have searched the catalogs from the NASA/IPAC Extragalactic Database (NED) and the SIMBAD Database for other potential counterparts within the 95\% error radius of the newly discovered $\gamma$-ray emission, but found no other plausible $\gamma$-ray sources (we have also examined coincident activities of nearby radio sources; see the Supplementary Material for the details).
Furthermore, we also checked whether SN 2017egm might coincidentally overlap with a $\gamma$-ray excess in the Fermi data with $TS > 25$ (which could originate from other sources). For this test, we estimated the occurrence rate of new sources with $TS > 25$ within any four-month interval based on real {\it Fermi-LAT} data (see Appendix~\ref{app:chance}).
The analysis shows that the probability of the observed $\gamma$-ray excess being due to a chance coincidence is very low ($\lesssim 4 \times 10^{-5}$). 
These results are in support of the $\gamma$-ray emission being true from SN 2017egm. In the above, the reported TS value of 30.1 is based on four-month data and a low energy threshold of 500 MeV of the data. We have tested that even if some other time ranges and thresholds are chosen, this signal remains robust (see Table~\ref{tab} of the Supplementary Materials).

The magnetar nebulae model of the SN 2017egm has been systematically discussed, and the self-consistent modeling predicted that the high-energy $\gamma$-ray emission would have a peak luminosity of $\sim 10^{43}~{\rm erg~s^{-1}}$ and hence may be detectable within a few months after the stellar explosion \cite{Vurm_2021,2023arXiv230206686A}. To compare the observational data with the predicted 0.1 $-$ 500 GeV light curve for a nebula magnetization of $\varepsilon_{B}=10^{-4}$ from Figure 7 of Ref.~\citep{Vurm_2021}, we performed data analysis in the 100 MeV to 500 GeV energy range to derive the light curve. The results are presented in Fig.~\ref{fig:fermi} (for comparisons with model curves of other nebula magnetizations, please see the Appendix~\ref{app:magnetization}). Clearly, both the rise time, the peak luminosity as well as the decline behavior of the GeV emission are consistent with the prediction of the magnetar nebulae model. Such consistency is in support of the millisecond magnetar central engine model of SN 2017egm.

We then suggest that SN 2017egm is a $\gamma$-ray emitter and the magnetars are the central engine of some superluminous supernovae, as initially suggested in \cite{2007ApJ...666.1069M,2010ApJ...717..245K,2010ApJ...719L.204W}. 
Nevertheless, due to the strong self-absorption of the GeV-TeV emission accompanying the superluminous supernovae, the detection of such signal is challenging for {\it Fermi}-LAT. GeV detections from SLSNe will be one of the important scientific objectives for future $\gamma$-ray telescopes, such as the Very Large Area $\gamma$-ray Space Telescope (VLAST). VLAST is a next-generation $\gamma$-ray detector under proposal, which will have a large acceptance of $\sim12\,\rm m^2\,sr$ in the GeV-TeV energy range \cite{2022AcASn..63...27F}. With this more sensitive instrument, GeV emission signals fainter than that of SN 2017egm will be detectable. In addition, future survey projects (e.g., the Legacy Survey of Space and Time at the Vera C. Rubin Observatory \cite{2019ApJ...873..111I}) will rapidly discover many more SLSNe. The combination of these upcoming facilities will provide a sample of GeV-detected SLSNe, the study of which will help us better understand the physical mechanisms (energy source, particle acceleration, magnetization, radiation process, etc) of SLSNe.

\section*{\label{sec::acknowledgments}Acknowledgments}


This work is supported in part by the National Key Research and Development Program of China (No. 2022YFF0503302), National Natural Science Foundation of China (No. 12588101 and No. 12103001), the New Cornerstone Science Foundation through the XPLORER PRIZE, and Anhui project (Z010118169).

%

\clearpage

\appendix

\setcounter{figure}{0}
\renewcommand\thefigure{S\arabic{figure}}
\setcounter{table}{0}
\renewcommand\thetable{S\arabic{table}}

\section*{Supplemental Material}
\section{Spectral analysis}
\begin{figure}[!ht]
\centering
\includegraphics[width=0.48\textwidth]{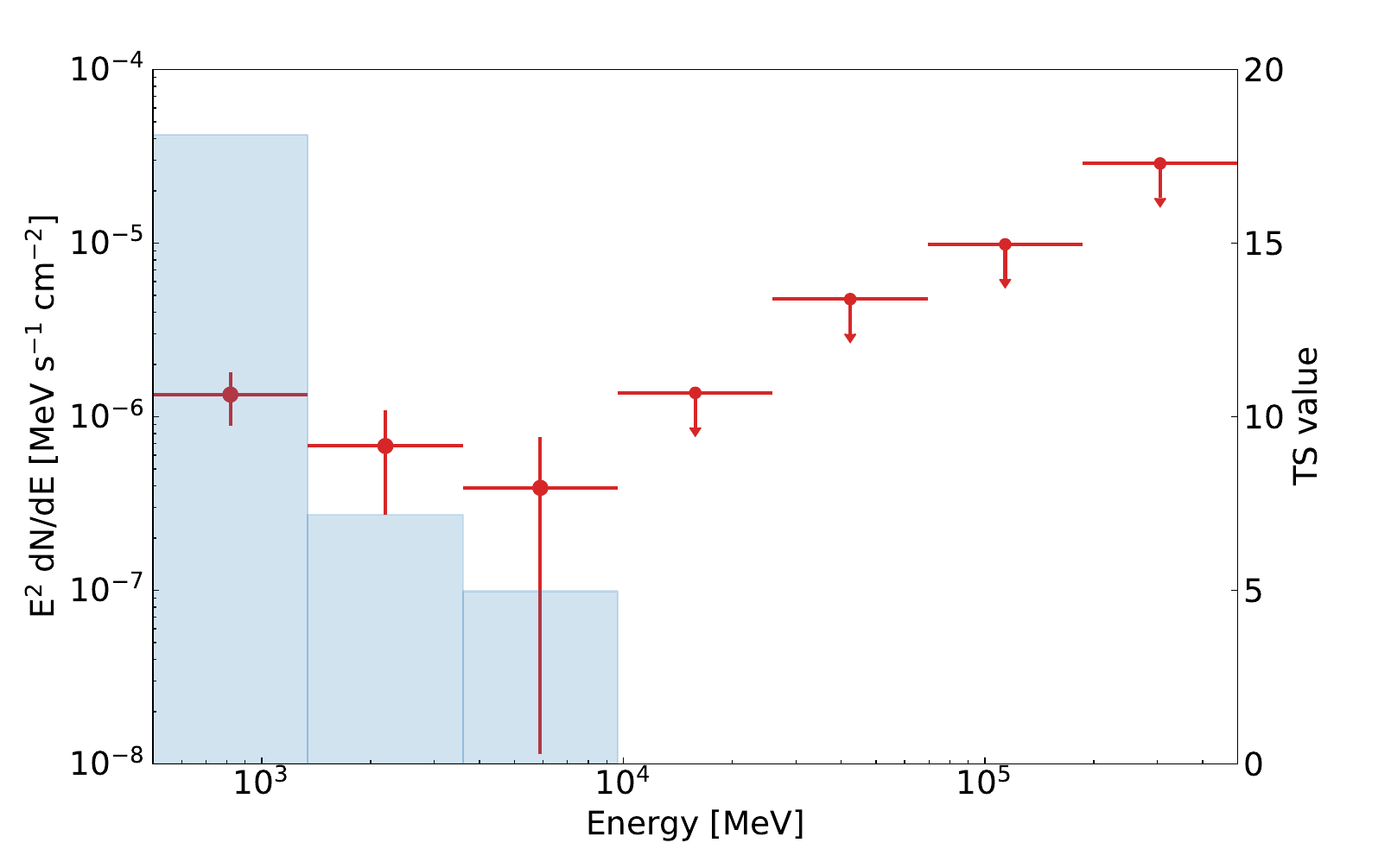}
\caption{Spectral energy distribution of the new $\gamma$-ray source.}
\label{fig:sed}
\end{figure}

We show in Fig.~\ref{fig:sed} the spectral energy distribution of the new $\gamma$-ray source for the data from 2017 July 23 to 2017 November 23.  The energy range from 500 MeV to 500 GeV is divided into seven equal logarithmic energy bins. The blue-shaded regions represent the TS values (right axis). For the energy bin with TS $<4$, the 95\% C.L. upper limit is presented. The power-law fit to the spectrum yields a spectral index of $2.6 \pm 0.4$.

\section{The robustness of signal in {\it Fermi}-LAT data analysis}

\begin{figure}[!ht]
\centering
\includegraphics[width=0.48\textwidth]{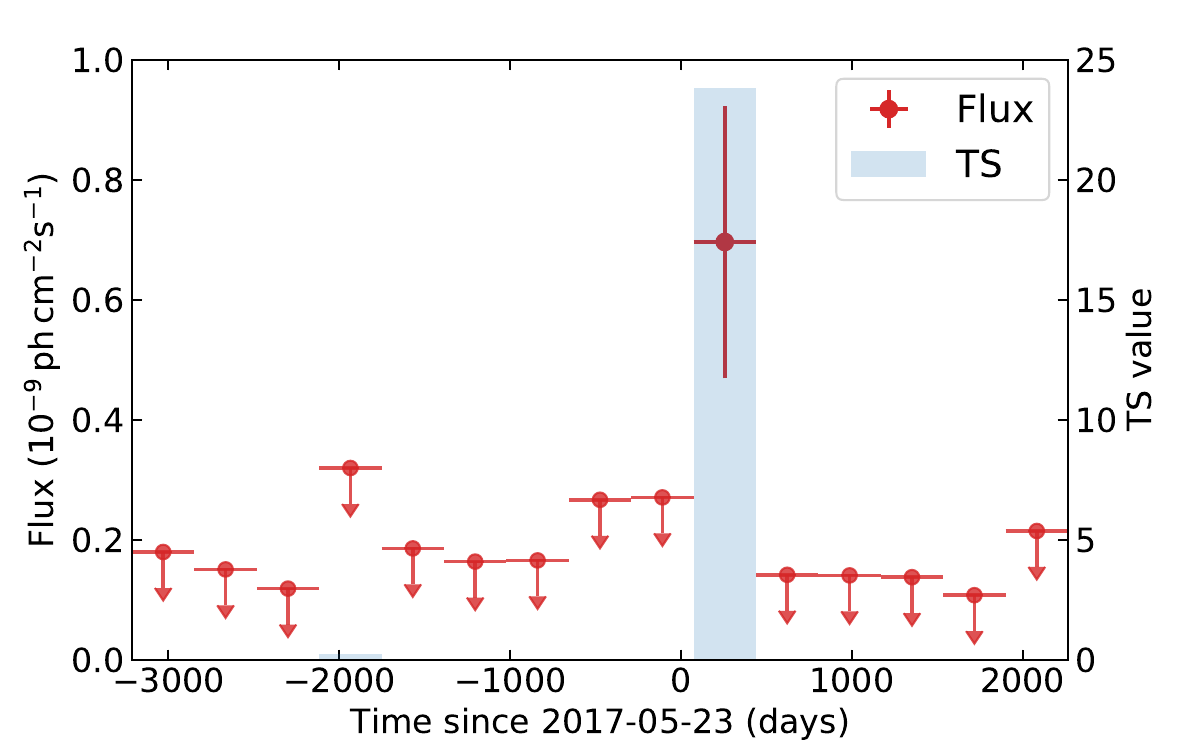}
\caption{Same as Fig.~\ref{fig:lc}, but for one-year bin of the $\gamma$-ray light curve for 15-year {\it Fermi}-LAT data.}
\label{fig:lcyear}
\end{figure}

In order to understand the behavior of the $\gamma$-ray emission in the direction of SN 2017egm. A 1-year bin $\gamma$-ray light curve is extracted. From Fig.~\ref{fig:lcyear}, we can find that the TS value of the 10th bin (from 2017 August 4 to 2018 August 4) is $\sim$ 23.8 and the TS values of the other bins are very small ($<$ 4), which is consistent with the Fig.~\ref{fig:tsmap}.

Further analysis indicated that the $\gamma$-ray emission is mainly concentrated in a period of 3 months, in the time range from 2017 July 23 to 2017 October 23. To further test the reliability of the $\gamma$-ray signal. We consider two different data selection situations, which include 1) varying the lower energy threshold (100MeV, 300MeV, 600MeV and 700MeV), and 2) setting $Z_{\rm max}$ to 90$^{\circ}$. The results of the binned likelihood analysis are summarized in Table \ref{tab}.

In addition, we also perform an unbinned likelihood analysis to test the $\gamma$-ray signal. The photons are selected from 500 MeV to 500 GeV within a 7$^{\circ}$ region of interest (ROI) centered at the SN 2017egm. The data cuts are similar to the binned analysis, including zenith angles less than 100$^{\circ}$ and (DATA\_QUAL ==1)\&\&(LAT\_CONFIG == 1). The results are also shown in Table \ref{tab}. For all the situations in Table \ref{tab}, the smallest TS value is 27.0 and the SN 2017egm are all within the 95\% C.L. error radius.

\begin{table*}[t]
\caption{The likelihood analysis results for SN 2017egm (3-month period).}
\begin{ruledtabular}
\begin{tabular}{lccccc}
\multicolumn{1}{c}{} &  TS  & (RA,DEC){\footnote{The $\gamma$-ray positions are determined by the {\it gtfindsrc} tool}} &R$_{\rm d}$\footnote{R$_{\rm d}$ is the angular seperation between the $\gamma$-ray position and the optical position of SN 2017egm}  & R$_{95\%}$ \footnote{R$_{95\%}$ is the 95\% error radius derived from the analysis.} \\
\multicolumn{1}{c}{} & &($^{\circ}$, $^{\circ}$)  &($^{\circ}$)&($^{\circ}$)\\
\hline
Binned, $Z_{\rm max}=100^{\circ}$, [0.1 - 500 GeV]  & 27.0 &(154.79, 46.39) & 0.06 &0.21\\
Binned, $Z_{\rm max}=100^{\circ}$, [0.3 - 500 GeV]  & 28.6 &(154.79, 46.40) & 0.06 &0.14 \\
Binned, $Z_{\rm max}=100^{\circ}$, [0.5 - 500 GeV]  & 38.4 &(154.80, 46.39)& 0.07&0.18\\
Binned, $Z_{\rm max}=100^{\circ}$, [0.6 - 500 GeV]  & 39.1 &(154.82, 46.39) & 0.07 &0.18 \\
Binned, $Z_{\rm max}=100^{\circ}$, [0.7- 500 GeV]  & 37.8 &(154.82, 46.39) & 0.07 &0.17 \\
Binned, $Z_{\rm max}=90^{\circ}$, [0.5 - 500 GeV]  & 31.7&(154.90, 46.33) &0.15&  0.20  \\
Unbinned, $Z_{\rm max}=100^{\circ}$, [0.5 - 500 GeV] & 32.5 &(154.80, 46.38)& 0.07&0.17\\

\end{tabular}
\end{ruledtabular}
\label{tab}
\end{table*}

To compare our results with that in Ref \cite{2023arXiv230206686A}, we also perform a binned likelihood analysis with event selection similar to the Ref \cite{2023arXiv230206686A}. We select photons with energies from 100 MeV to 500 GeV and consider the same period from 2017 May 23 to 2020 August 21. To reduce the contamination from the Earth limb, only the events with zenith angle less than $90^\circ$ are selected. The data are split into 200 $\times$ 200 spatial bins with 0.1$^{\circ}$ per pixel. A power-law spectrum ($dN/dE \propto E^{-\Gamma}$) with $\Gamma =$ 2.0 is applied to fit the SN 2017egm. 
The model includes all 4FGL-DR4 sources within 20$^{\circ}$ around the target, together with the diffuse emission templates ({\tt gll\_iem\_v07.fits} and {\tt iso\_P8R3\_SOURCE\_V3\_v1.txt}). 

First, we perform a fit of the entire data from 2017 May 23 to 2020 August 21. The TS value is $\sim$ 4.6 and the energy flux upper limit is $1.1 \times10^{-6}~{\rm MeV~cm^{-2}~s^{-1}}$ for the energy range from 100 MeV to 500 GeV. 
Then we split the data into seven time bins, the first bin is the first 90 days after the explosion and the other date is split into six bins ($\sim$ 6-month of each bin). For the first bin, the energy flux upper limit is $3.6 \times 10^{-6}~{\rm MeV~cm^{-2}~s^{-1}}$ with a TS $\sim$ 0.4. The TS value of the first 6-month bin is $\sim$ 9.5, the corresponding energy flux upper limit is $4.6 \times10^{-6} ~{\rm MeV ~cm^{-2}~s^{-1}}$. The peak TS value of the other 6-month bins is $\sim$ 2.0. These results are coincident with the Ref \cite{2023arXiv230206686A}.

\section{The lack of simultaneous activities of nearby  sources}
We searched for possible variable or burst activities from the nearby radio sources in the region of the $\gamma$-ray source. There is no strong evidence of simultaneous activities from nearby radio sources in the surveys conducted by the {GAIA}, {ZTF}, and {Chandra} satellites.

In our search for other potential contributions from burst or flare events in the vicinity of the $\gamma$-ray source using the GAIA survey, we found no significant luminous optical burst or flare in the {\it GAIA Data Release 3} catalog \cite{2023A&A...674A...1G}, except for SN 2017egm. The {\it GAIA Data Release 3} covers the survey period from 2017-01-01 to 2018-10-15. While there are two quasar candidates with magnitudes of $\sim 20\rm \, mag$ in the GAIA/G band photometry in the region, the amplitudes of their photometric light curves do not exceed $\sim 1\rm \, mag$. Therefore, the likelihood of these two quasar candidates (GAIA IDs: 810264031570565632, 810263378735530880) causing a burst event is very low.

We cross-matched the sources from the radio survey catalogs \cite{2018MNRAS.474.5008D,2021ApJ...914...42B,2011ApJ...737...45O,2015ApJ...801...26H,2010A&A...511A..53V,2017A&A...598A..78I,2002ApJS..143....1M,2023ApJS..267...37G,2021ApJS..255...30G,2021A&A...655A..17S,1998AJ....115.1693C} with the ZTF Data Release 20\footnote{ZTF data release: \url{https://www.ztf.caltech.edu/ztf-public-releases.html}} in a circular region with a radius of 0.2$^{\circ}$ located at R.A. 154.79$^{\circ}$ and DEC. 46.40$^{\circ}$. Simultaneously, we searched and cross-matched the radio sources in the same region with the five Chandra X-ray exposures (Obsid: 19034, 19035, 19036, 19033, PI: Margutti; Obsid: 22501, PI: Lehmer). We combined the five Chandra observations using the script $merge\_obs$\footnote{$merge\_obs$: \url{https://cxc.harvard.edu/ciao/threads/merge_all/}} from the Chandra standard software CIAO. Three X-ray sources were detected to have a significance $>5\sigma$ with the CIAO script $wavdetect$\footnote{$wavdetect$: \url{https://cxc.cfa.harvard.edu/ciao/ahelp/wavdetect.html}} in the merged observation and were cross-matched with radio sources within a radius of $9.0^{\prime\prime}$. Two of the three X-ray sources were cross-matched with the ZTF optical catalogue. The features of the three X-ray sources are listed in Table \ref{tab2}.

\begin{table*}
\centering
\caption{The features of the three cross-matched Chandra X-ray sources in the direction of the GeV $\gamma$-ray transient. }
\begin{tabular}{cccccccc}%
\hline
\hline
\multicolumn{1}{c}{}  cross-match & \# of sources & band & position & offset to radio obj & flux density & burst event? \\
\multicolumn{1}{c}{}  & cross-matched &  & (RA, DEC)  & ($^{\prime\prime}$) & ($\rm erg/s/cm^2/Hz$) &  (light curve) \\
\hline\\
\multicolumn{7}{l}{ Radio Source 1 (154.77099, 46.45466): NVSS, $f_{\nu, \rm 1.4GHz}=1.0\times 10^{-25}\, \rm erg/s/cm^2/Hz$, Ref\cite{2018MNRAS.474.5008D} } \\
\\
Chandra &  1/1  &  0.5-7.0 keV &  154.76867, 46.45439 & $8.4^{\prime\prime}$  & $2.9\pm 4.9\times 10^{-32}$  & no \\
ZTF  &  1/1  &  g & 154.77141, 46.45413 & $2.4^{\prime\prime}$ & $5.79^{+6.39}_{-1.47} \times 10^{-27}$   & no \\
ZTF  &  1/1  &  r & 154.77141, 46.45413 & $2.4^{\prime\prime}$ &  $1.12^{+1.34}_{-0.28} \times 10^{-26}$   & no  \\
ZTF  &  1/1  &  i & 154.77141, 46.45413 & $2.4^{\prime\prime}$ &  $1.46^{+3.85}_{-2.95} \times 10^{-26}$   & no  \\
\hline\\
\multicolumn{7}{l}{ Radio Source 2 (154.74294, 46.45402): NVSS, $f_{\nu , \rm 1.4GHz}=3.7\times 10^{-26}\, \rm erg/s/cm^2/Hz$, Ref\cite{2011ApJ...737...45O}} \\
\\
Chandra &  1/1  &  0.5-7.0 keV &  154.74325, 46.45343  &  $2.3^{\prime\prime}$  &  $3.95\pm 7.0 \times 10^{-32}$ &  no \\
ZTF  &  1/3  &  g & 154.74283, 46.45441  & $1.4^{\prime\prime}$  & $2.60 ^{+1.08}_{-0.89}\times 10^{-27}$  & no \\
ZTF  &  1/3  &  r &  154.74283, 46.45441 & $1.4^{\prime\prime}$   & $4.38 ^{+1.68}_{-1.71}\times 10^{-27}$  & no  \\
ZTF  &  2/3  &  r &  154.74167, 46.45414 & $4.6^{\prime\prime}$   & $8.02 ^{+7.77}_{-3.02}\times 10^{-27}$  & no  \\
ZTF  &  2/3  &  i &  154.74167, 46.45414 & $4.6^{\prime\prime}$   & $1.02 ^{+0.37}_{-0.16}\times 10^{-26}$  & no  \\
ZTF  &  3/3  &  g &  154.74360, 46.45497 & $4.2^{\prime\prime}$   & $1.41 ^{+1.12}_{-0.56}\times 10^{-27}$  & no  \\
\hline\\
\multicolumn{7}{l}{ Radio Source 3 (154.56401, 46.44057): FIRST, $f_{\nu , \rm 1.4GHz}=1.3\times 10^{-26}\, \rm erg/s/cm^2/Hz$, Ref\cite{2002ApJS..143....1M}} \\
\\
Chandra &  1/1  &  0.5-7.0 keV &  154.5642600, 46.4411149  &  $2.2^{\prime\prime}$  & $7.36\pm 12.4\times 10^{-32}$ & no \\
ZTF  &  0/0  &  - & - &  - & - & - \\
\hline
\hline
\end{tabular}
\label{tab2}
\end{table*}

We also considered the presence of strong jets in the cross-matched sources when assuming them to be quasars. We calculated the radio loudness $R=\frac{f_{\nu , 1.4\rm GHz}}{f_{\nu , \rm ZTF-g}}$ using the flux density values in Table \ref{tab2}. The effective wavelength of the ZTF-g band is $472.2\, \rm nm$. The radio loudness of these ZTF sources in the $\gamma$-ray region falls between 'radio-quiet' and 'radio-loud' quasars based on their radio and ZTF-g band emissions. Their radio loudness range is $R\sim 14-26$, which is lower than that of 'radio-loud' quasars ($R\sim100$, \cite{2006ApJ...637..669L}). The low radio loudness indicates that these assumed quasar candidates are unlikely to have strong jets.

To verify if the X-ray sources are associated with burst events, we assessed the flux variability of each source using the CIAO script $srcflux$\footnote{$srcflux$: \url{https://cxc.cfa.harvard.edu/ciao/ahelp/srcflux.html}}. The three sources exhibit faint X-ray emission ($1\times 10^{-15}\sim 3\times 10^{-14}\, \rm erg\, s^{-1}\, cm^{-2}$) in the Chandra X-ray observations. We did not observe any obvious burst events in the X-ray light curves of the three X-ray sources.

\section{The probability of chance coincidence with a background $\gamma$-ray excess}
\label{app:chance}

Before claiming a detection, it is necessary to estimate the probability of chance coincidence (e.g., \cite{Prokhorov:2022dpl}). 
To examine whether SN 2017egm might coincidentally overlap with a $\gamma$-ray excess with $TS > 25$ which may originate from other sources not related to SN 2017egm, we estimate the occurrence rate of new sources with $TS > 25$ within any four-month interval.

We randomly select 24 regions of size $14^\circ \times 14^\circ$ in the high Galactic latitude area ($|b| > 30^\circ$, since SN 2017egm is at a Galactic lattitude of $b=54.4^\circ$). The distance between the centers of any two regions is required to be larger than $20^\circ$ to ensure that the analysis regions do not overlap. For each region, the 15 years of Fermi data are divided into 45 equal time intervals with a duration of four months each. We perform binned likelihood analysis for each time interval in each region and generate residual TS maps, resulting in a total of $24 \times 45 = 1080$ TS maps. All analyses use the same procedure as in the main text for SN 2017egm (same energy range, data selection, etc.). 

Among all the TS maps, only 5 excesses exhibit maximum TS values greater than 25. The corresponding occurrence rate is ${5}/{(1080 \times 0.0597)} \sim 0.0775\ \text{sr}^{-1}$, where 0.0597 is the solid angle of a $14^\circ \times 14^\circ$ region.
Therefore, the chance probability of an excess with TS $> 25$ appearing within a radius of 0.2$^\circ$ (both the localization uncertainty and the angular deviation of the $\gamma$-ray signal are smaller than 0.2$^\circ$) is $\sim 9\times10^{-7}$.
Here we have assumed that the 45 time intervals in the same region are statistically independent, though in reality they may not be completely independent. If, conservatively, only considering 24 independent regions, then the occurrence rate is ${5}/{(24 \times 0.0597)} \sim 3.49\ \text{sr}^{-1}$, corresponding to a chance probability of $\sim 4\times10^{-5}$. These results indicates the probability of the observed $\gamma$-ray excess being due to a chance coincidence is very low.

\begin{figure}
\centering
\includegraphics[width=0.5\textwidth]{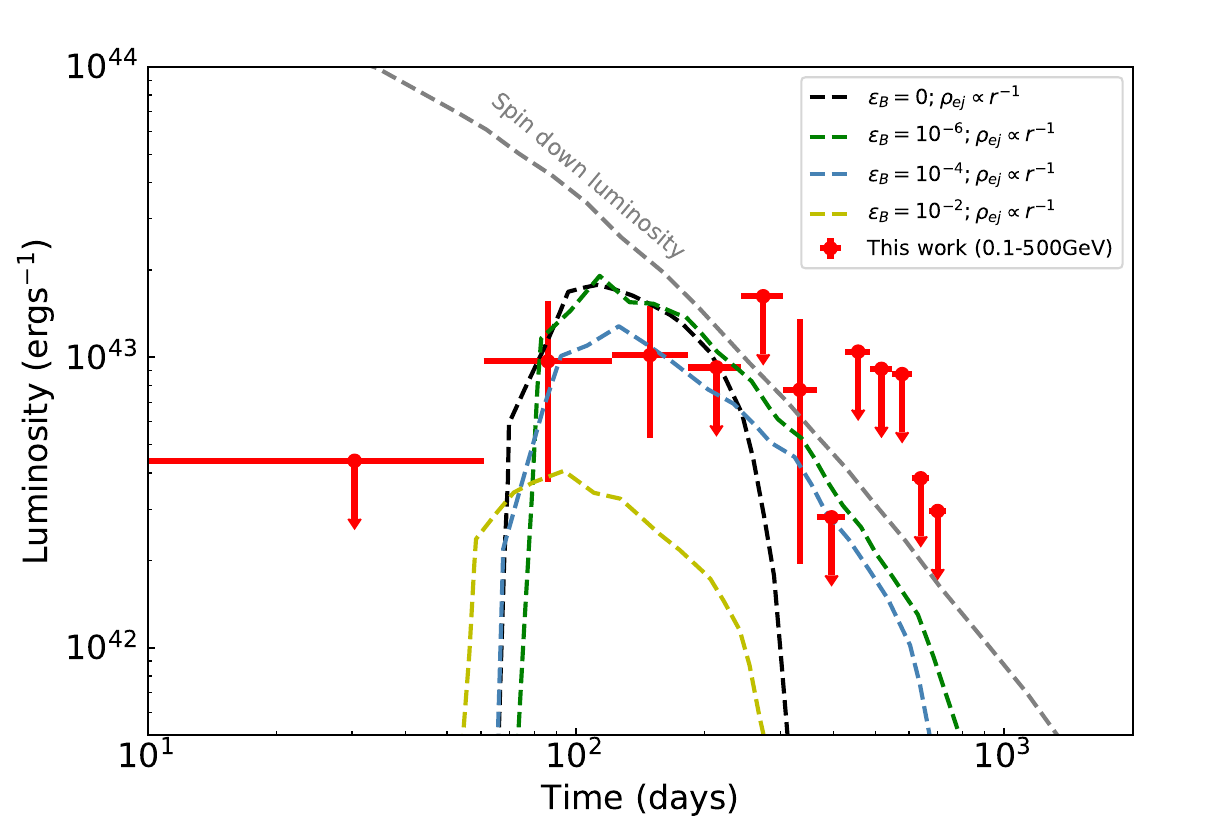}
\caption{Same as Fig.~\ref{fig:fermi}, but showing the four theoretical curves corresponding to different values of the nebula magnetization $\varepsilon_B = 0, 10^{-6}, 10^{-4} ,10^{-2}$ \cite{Vurm_2021}.}
\label{fig:b}
\end{figure}

\section{Model uncertainty caused by nebula magnetization}
\label{app:magnetization}
As stated in Ref.~\cite{Vurm_2021}, after choosing the engine and ejecta properties to match the optical emission near the light-curve peak, the quantity that has the greatest impact on the gamma-ray emission is the nebula magnetization $\varepsilon_B$. 
When the magnetization is high, the energy leaves the nebula in synchrotron radiation, which is absorbed and gets thermalized within the ejecta. Only when the magnetization is low does the nebula radiates the engine energy predominantly via inverse Compton emission, which will not be thermalized and can then be observed. Overall, our results favor a relatively low magnetization of $\varepsilon_B\le10^{-4}$, but values of $\varepsilon_B=10^{-4}$, $10^{-6}$ and 0 can all generally agree with the observational data with $\varepsilon_B\sim10^{-4}$ giving the best match. 
In Fig. \ref{fig:b}, we compare the model spectra expected for different nebula magnetization with the observations.

\end{document}